\documentclass[useAMS,usenatbib]{mn2e}

\usepackage{amsmath}
\usepackage{amstext}
\usepackage{amsbsy}
\usepackage{amsopn}
\usepackage{graphicx}
\usepackage{amssymb}
\usepackage{subfigure}

\def\be{\begin{equation}}
\def\ee{\end{equation}}

\def\kms{\,{km\,s$^{-1}$}}

\def\cmd{\,{\rm {cm^{-2}}}}

\def\LJ{$L_{\rm J}$}

\def\Lya{Ly$\alpha$~}
\def\Lyb{Ly$\beta$~}

\def\NHI{{$N($\HI$)$}}

\def\loss{{lines of sight}}
\def\los{{line of sight}}
\newcommand{\HI}{\mbox{H\,{\sc i}}}

\title[]{Tracing the  gas at redshift 1.7-3.5  with the Lyman-$\alpha$
  forest: the FLO approach  \thanks{Based on observations collected at
  the  European  Southern  Observatory  Very  Large  Telescope,  Cerro
  Paranal,  Chile -- Programs  166.A-0106(A) and  during commissioning
  and science verification of UVES }}

\author[F. Saitta et al.]{F. Saitta$^{1,2,3}$, V. D'Odorico$^{1}$, M. Bruscoli$^{1,4}$, S.Cristiani$^{1,5}$,
  P. Monaco$^{1,3}$, M. Viel$^{1,5}$\\
$^1$INAF-Osservatorio Astronomico di Trieste, via Tiepolo 11, I-34143 Trieste, Italy\\
$^2$European Southern Observatory, Karl-Schwarzschil-Str. 2, D-85748 Garching, Germany \\
$^3$Dipartimento di Astronomia, Universit\`a di Trieste, via Tiepolo 11, I-34143 Trieste, Italy \\
$^4$INAF-IRA, L.go E. Fermi 5, I-50125 Firenze, Italy\\
$^5$INFN/National Institute for Nuclear Physics, Via Valerio 2, I-34127 Trieste, Italy\\}

\begin{document}


\pagerange{\pageref{firstpage}--\pageref{lastpage}} \pubyear{2007}

\maketitle

\label{firstpage}

\begin{abstract}
We  present  {\small  FLO}  (From  Lines  to  Over-densities),  a  new
technique  to reconstruct  the  hydrogen density  field  for the  \Lya
forest lines observed  in high resolution QSO spectra.   The method is
based  on the  hypothesis that  the  \Lya lines  arise in  the low  to
intermediate density  intergalactic gas and  that the Jeans  length is
the typical  size of the  \Lya absorbers.  The reliability  of {\small
FLO}  is  tested  against  mock  spectra  obtained  from  cosmological
simulations.  The  recovering algorithm gives  satisfactory results in
the range  from the  mean density to  over-densities of $\sim  30$ and
reproduces correctly the correlation function of the density field and
the  1D power spectrum  on scales  between $\sim  20$ and  60 comoving
Mpc. A sample  of \Lya forests from 22 high  resolution QSO spectra is
analysed, covering  the redshift range  $1.7\la z \la 3.5$.   For each
line of sight, we fit Voigt  profiles to the lines of the \Lya forest,
providing the  largest, homogeneous sample  of fitted \Lya  lines ever
studied.  The  line number density  evolution with redshift  follows a
power-law   relation:   $dn/dz=(166\pm  4)\,[(1+z)/3.5]^{(2.8\pm0.2)}$
(1\,$\sigma$  errors).  The  two-point correlation  function  of lines
shows a signal up to separations  of $\sim 2$ comoving Mpc; weak lines
($\log N($\HI$)  < 13.8$)  also show a  significant clustering  but on
smaller scales ($r  \la 1.5$ comoving Mpc).  We  estimate with {\small
FLO} the hydrogen density field toward the 22 observed lines of sight.
The redshift  distribution of the average densities  computed for each
QSO  is  consistent with  the  cosmic  mean  hydrogen density  in  the
analysed redshift  range.  The two-point correlation  function and the
1D power spectrum of the  $\delta$ field are estimated.  They are both
consistent  with the analogous  results computed  from hydro-simulated
spectra  obtained in  the  framework of  the concordance  cosmological
model. The correlation function shows clustering signal up to $\sim 4$
comoving Mpc.
\end{abstract}


\section{Introduction}
\label{intro}

Semi-analytical         and         hydro-dynamical        simulations
\citep[e.g.,][]{cen94,zhang95,hernquist96,
miralda96,bi97,dave97,zhang97,theuns98,machacek00}  suggest  that  the
\Lya forest arises from  fluctuations of the low density intergalactic
medium  (IGM) that  trace  the underlying  matter  density field  over
cosmic time.  The  dynamical state of the low  density IGM is governed
mainly by the Hubble expansion and by gravitational instabilities.  As
a  consequence,  the  physics  involved  is quite  simple  and  mildly
non-linear.   The statistical  analysis  of the  \Lya forest  provides
information on the dynamical growth  and thermal state of the IGM, and
on the  correlation properties of  the (dark) matter in  the Universe.
Correlations  of   the  \Lya  forest   lines  were  detected   with  a
$4-5\,\sigma$ confidence by various  authors at typical scales $\Delta
v  \la 350$  \kms\ observing  at high  resolution individual  lines of
sight  (\citealt{cristiani95}  and   \citealt{lu96}  at  $z\sim  3.7$;
\citealt{cristiani97} at $z\sim3$; \citealt{kim01} at $z\sim2$).  This
velocity range  corresponds to scales $\la 2.5\  h^{-1}$ Mpc (assuming
negligible   peculiar  velocities).    The  ``cosmic   web''  scenario
\citep{bond}  is favoured  against that  of a  population  of pressure
confined clouds  \citep{sargent80} thanks also to the  analysis of the
line  correlation observed  in  close  pairs of  QSO  lines of  sight,
implying     absorber     sizes    of     a     few    hundred     kpc
\citep[e.g.,][]{smette92,smette95,bechtold94,fang96,dinshaw97,crottsfang98,vale98,petitjean98,rauch01,young01,becker04}.
The analysis of multiple lines of sight at slightly larger separations
(smaller  than a  few arcminutes),  makes it  possible to  compute the
transverse  correlation  function for  which  a  clustering signal  is
detected up to  velocity separations of $\sim 200$  \kms, or about $3\
h^{-1}$ comoving Mpc \citep{roll03,tomography,copp06}.

Traditionally, absorption spectra  were decomposed into Voigt profiles
which  were  then   identified  with  individual  discrete  absorption
systems.  Information on the physical state of the gas originating the
absorptions comes  directly from the fit  parameters: redshift, column
density and Doppler broadening (linked to the temperature). In the new
paradigm  the emphasis  of  the analysis  has  shifted to  statistical
measures of the  transmitted flux (e.g. the flux  power spectrum) more
suitable  for  absorption arising  from  a  continuous density  field.
However the interpretation of statistical quantities of the continuous
flux field and their relation  with the physical properties of the gas
requires   a   non-trivial   comparison  with   full   hydro-dynamical
high-resolution simulations that are computationally expensive.

The  aim of  this paper  is  to extend  the line  fitting approach  by
identifying  a  new  statistical  estimator  linked  to  the  physical
properties  of  the underlying  IGM.   This  new  estimator will  also
overcome  the  two  main   drawbacks  of  the  Voigt  fitting  method:
\par\noindent   (i)  the  subjectivity   of  the   decomposition  into
components:  the   same  absorption  can  be   resolved  by  different
scientists (or software  tools) in different ways, both  in the number
of components, and in the values of the output parameters for a single
component;  \par\noindent (ii)  the blanketing  effect of  weak lines:
they can be  hidden by the stronger lines, so  that their exact number
density is unknown and has  to be inferred from statistical arguments.
Unfortunately, since  the weak lines  are also the most  numerous, the
uncertainty  in their exact  number is  transformed into  a systematic
error of the computed statistical quantities.

This  new  estimator is  identified  in  the  hydrogen density  field,
$n_{\rm H}$,  which is  linked to the  measured \HI\  column densities
through the formula \citep{Sch}:
\begin{eqnarray}
\label{tra}
&   N({\rm  HI})   \simeq 3.7 \times 10^{13}\ \cmd
  (1+\delta)^{1.5-0.26\alpha} T_{0,4}^{-0.26}  \Gamma_{12}^{-1} \\
& \times \left(\frac{1+z}{4}\right)^{9/2} 
\left(\frac{\Omega_{\rm b}\,h^2}{0.024}\right)^{3/2}
\left(\frac{f_g}{0.178}\right)^{1/2},\nonumber
\end{eqnarray}
\noindent
where, $\delta  \equiv n_{\rm H}/\langle  n_{\rm H}\rangle -1$  is the
density contrast,  $T_{0,4} \equiv T_0  / 10^4$ is the  temperature at
the  mean  density,  $\Gamma_{12}  \equiv \Gamma/10^{-12}$  is  the  H
photo-ionisation rate, $f_g  \approx \Omega_{\rm b}/\Omega_{\rm m}$ is
the fraction of the mass in gas and $\alpha$ depends on the ionisation
history  of the  Universe.   Equation~\ref{tra} relies  on three  main
hypotheses:  (i)  \Lya  absorbers   are  close  to  local  hydrostatic
equilibrium, i.e.  their characteristic  size will be typically of the
order  of  the   local  Jeans  length  (\LJ);  (ii)   the  gas  is  in
photo-ionisation   equilibrium;   (iii)   the   equation   of   state,
$T=T_0\,(\delta +  1)^{\alpha}$ holds for  the optically thin  IGM gas
\citep{huignedin}.

The procedure  to recover the  H density field  from the list  of \Lya
line column densities in a QSO  line of sight, has been dubbed {\small
FLO} (From Lines to Over-densities).

The paper  is organised as  follows: Section~2 describes  the observed
data  sample  used  for  our  analysis and  presents  the  statistical
measures obtained for the  fitted \Lya lines; Section~3 introduces the
hydrogen density  field as a statistical estimator,  and describes the
construction algorithm;  Section~4 presents the  simulated spectra and
the test of reliability of  the method with this dataset; in Section~5
the new algorithm is applied  to the observed data sample; finally, we
draw our conclusions in Section~6.

The cosmological model adopted  throughout this paper corresponds to a
`fiducial'   $\Lambda$CDM   Universe   with  parameters,   at   $z=0$,
$\Omega_{\rm    m}=0.26,\   \Omega_{\Lambda}=0.74,\    \Omega_{\rm   b
}=0.0463$, $n_s=0.95$,  $\sigma_8 = 0.85$  and $H_0 = 72$  km s$^{-1}$
Mpc$^{-1}$ \citep[the B2 set of parameters of][]{viel04}.

\section{Observed data sample}

\begin{table}
\centering
\begin{tabular}{lclc}
\hline                      
QSO&$z_{\rm em}$&J mag& \Lya $\Delta_{z}$\\          
\hline
HE1341-1020 & 2.139& 18.68 &1.658-2.087\\
Q0122-380   & 2.203& 17.34 &1.711-2.150\\
PKS1448-232 & 2.220& 17.09 &1.725-2.166\\
PKS0237-23  & 2.233& 16.61 &1.737-2.179\\
J2233-606   & 2.250& 16.97 &1.753-2.196\\
HE0001-2340 & 2.267& 16.74 &1.765-2.213\\
HE1122-1648 & 2.400& 16.61 &1.878-2.344\\
Q0109-3518  & 2.407& 16.72 &1.884-2.350\\
HE2217-2818 & 2.415& 16.47 &1.891-2.358\\
Q0329-385   & 2.440& 17.20 &1.912-2.383\\
HE1158-1843 & 2.451& 17.09 &1.921-2.394\\
HE1347-2457 & 2.560& 17.35 &2.016-2.502\\
Q0453-423   & 2.662& 17.69 &2.100-2.602\\
PKS0329-255 & 2.698& 17.88 &2.125-2.636\\
HE0151-4326 & 2.761& 17.48 &2.186-2.699\\
Q0002-422   & 2.768& 17.50 &2.189-2.705\\
HE2347-4342 & 2.878& 17.12 &2.281-2.814\\
HS1946+7658 & 3.051& 16.64 &2.429-2.984\\
HE0940-1050 & 3.088& 17.08 &2.463-3.020\\
Q0420-388   & 3.122& 17.44 &2.489-3.053\\
PKS2126-158 & 3.275& 17.54 &2.620-3.204\\
B1422+231   & 3.620& 16.22 &2.911-3.543\\
\hline
\end{tabular}
\caption{Summary  of  the  main  characteristics of  our  QSO  sample.
(see text).}
\label{tab1}
\end{table}
\begin{figure}
\begin{center}
\includegraphics[width=8cm]{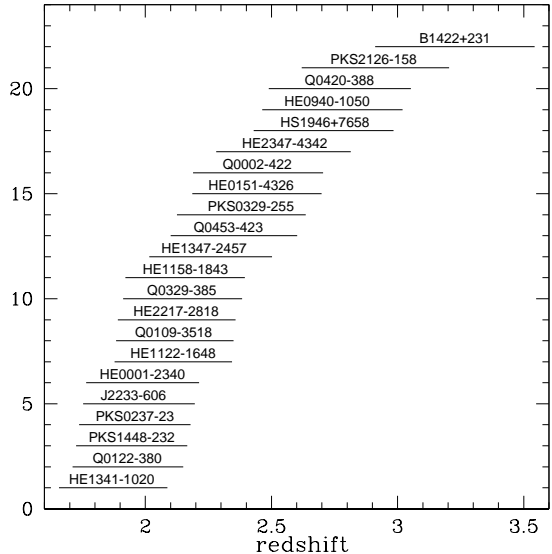}
\caption{\Lya forest redshift coverage  of the QSOs in our 
  sample. }
\label{qsodist}
\end{center}
\end{figure}
Most of  the observational data used  in this work  were obtained with
the UVES spectrograph \citep{dekker} at the Kueyen unit of the ESO VLT
(Cerro Paranal,  Chile) in  the framework of  the ESO  Large Programme
(LP): ``The Cosmic Evolution of the IGM'' \citep{bergeron04}.  Spectra
of 18 QSOs were obtained in  service mode with the aim of studying the
physics of the IGM in the  redshift range 1.7-3.5.  The spectra have a
resolution $R \sim 45000$ and a typical signal to noise ratio (SNR) of
$\sim  35$  and 70  per  pixel at  3500  and  6000 \AA,  respectively.
Details    of    the    data     reduction    can    be    found    in
\citet{chand04,aracil04}.  \\ We  added to the main sample  4 more QSO
spectra with comparable resolution and SNR:
\par\noindent  -  J2233-606 \citep{j2233}.  Data  for  this QSO  were
acquired during the commissioning of UVES in October 1999.
\par\noindent  - HE1122-1648  \citep{kim02}.  Data  for this  QSO were
acquired  during   the  science  verification  of   UVES  in  February
2000. 
The reduced and  fitted spectrum was kindly provided  to us by Tae-Sun
Kim.
\par\noindent -  HS1946+7658 \citep{kirk97}.   Data for this  QSO were
acquired with  Keck/HIRES in  July 1994. 
%
\par\noindent  - B1422+231  \citep{rauch96}.  Data for  this QSO  were
acquired with Keck/HIRES  in 1996. 
The reduced and  fitted spectrum was kindly provided  to us by Tae-Sun
Kim.

\begin{figure}
\begin{center}
\includegraphics[width=8 cm]{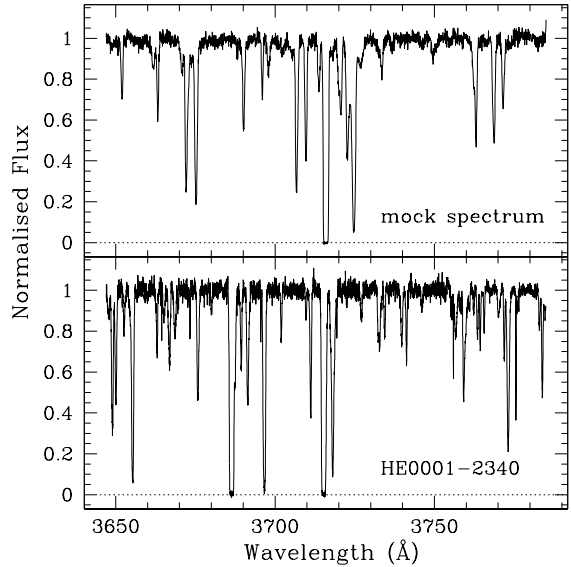}
\caption{ Portion of  the \Lya forest for an  observed (lower panel)
and a simulated (upper panel) line of sight in our sample.
}
\label{spectra}
\end{center}
\end{figure}

Table~\ref{tab1}  summarises the  main properties  of our  QSO sample.
None of our QSOs is a Broad Absorption Line (BAL) object.
Magnitudes   are  taken  from   the  GSC-II   catalogue  \citep{GSC2}.
Figure~\ref{qsodist} shows  the distribution  in redshift of  the \Lya
forests for  all the QSOs of  the sample.  We considered  for each QSO
the redshift range  between 1000 \kms\ red-ward of  the \Lyb emission,
in  order to avoid  contamination by  associated Ly$\beta$  lines, and
5000  \kms\ blue-ward  of  the  \Lya emission  to  exclude the  region
affected by the proximity effect due  to the ionising flux of the QSO.
The coverage  is good over the whole  redshift range $z\simeq1.7-3.5$,
with most  of the  signal concentrated between  $z\sim 2$ and  2.5. In
Fig.~\ref{spectra}, we  show a portion of  the \Lya forest  of the QSO
HE0001-2340  compared  with  the  same  wavelength region  in  a  mock
spectrum extracted  from the considered  simulation box at  $z=2$ (see
section~\ref{sim}).

\subsection{Creation of the line lists}
\label{subsec:line}

All the lines  in the \Lya regions of the LP  QSOs plus J2233-606 were
fitted  with  the  {\small  FITLYMAN} tool  \citep{font:ball}  of  the
{\small          ESO           MIDAS}          data          reduction
package\footnote{http://www.eso.org/midas}.   In the  case  of complex
saturated  lines we  used the  minimum number  of components  to reach
$\chi^{2} \le 1.5$.   Whenever possible, the other lines  in the Lyman
series  were  used  to  constrain   the  fit. The  spectra  of
HE1122-1648,  HS1946+7658, B1422+231  and all  the simulated  lines of
sight        were         fitted        with        the        {\small
VPFIT}\footnote{http://www.ast.cam.ac.uk/$\sim$rfc/vpfit.html}
package. Both  software tools model  absorption features with  a Voigt
profile convolved with the instrument line spread function.
The minimum \HI\ column density detectable at 3\,$\sigma$, at the
lowest SNR  of the spectra in  our sample, is  $\log N($\HI$)\simeq 12$
cm$^{-2}$.

Metals in  the forest were  identified and the  corresponding spectral
regions  were  masked  to   avoid  effects  of  line  blanketing.   We
eliminated \Lya lines with Doppler  parameters $b \le 10$ km s$^{-1}$,
that are likely unidentified metal absorptions. In a total amount
of 8435 fitted \Lya lines,  368 (4.4\%) fall in the masked intervals,
1150 (13.6\%) are at less  than 1000 \kms\ red-ward the \Lyb emission
or at  less than  5000 \kms\ blue-ward  the \Lya emission,  while 599
were  eliminated because they have  $b  \le  10$  (7.1\%).
The output of this analysis is a  list of \Lya lines for each QSO with
central redshift, \HI\ column density and Doppler parameter.

In the  line fitting approach  to the study  of the \Lya  forest, each
line is considered  as the signature of an  absorber. As a consequence
statistical measures  are computed  with the population  of absorption
lines, representative  of the population of absorbers.   Our sample of
fitted \Lya lines is the  largest, homogeneous sample ever gathered up
to now.  We  will use it to compute the  number density evolution with
redshift and the two point correlation function of lines.

\subsection{Line Number density evolution}
\label{dn_dz}

\begin{figure}
\begin{center}
\includegraphics[width=9cm]{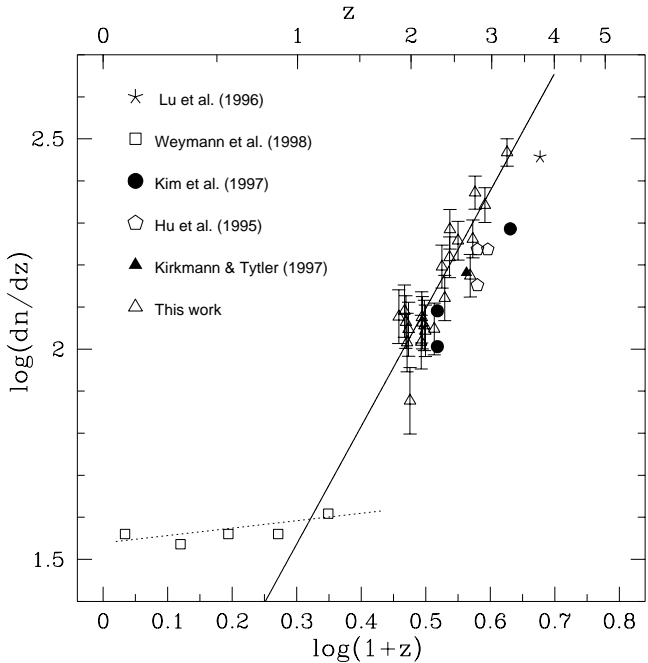}
\caption{Number density  evolution of the  \Lya forest lines  over the
column density range $13.64 < \log N($\HI$) < 17$ cm$^{-2}$ for the 22
QSOs in our  sample (open triangles).  The solid  line traces the best
linear fit  obtained for  those data (see  text).  For  comparison, we
report also previous  measurements at high redshift and  the result of
the low redshift HST campaign.}
\label{dndz}
\end{center}
\end{figure}

\begin{figure}
\begin{center}
\includegraphics[width=9cm]{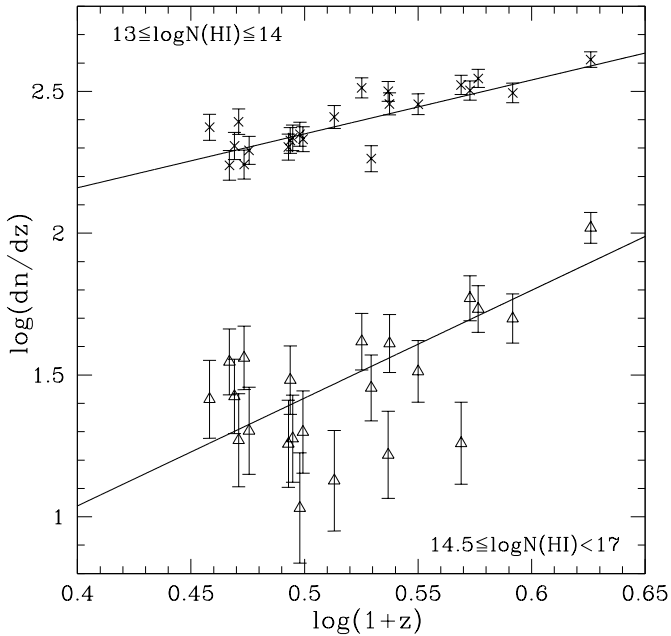}
\caption{Number density  evolution of the  \Lya forest lines  over the
two column density ranges $13  \le \log N($\HI$) \le 14$ (crosses) and
$14.5 \le  \log N($\HI$) < 17$  cm$^{-2}$ (open triangles)  for the 22
QSOs in  our sample.  The lines are  the best linear fits  for the two
distributions (see text).}
\label{dndz2}
\end{center}
\end{figure}

The line number density per unit of redshift is generally approximated
as $dn/dz  = (dn/dz)_0 (1+z)^{\beta}$, where $(dn/dz)_0$  is the local
comoving line  number density of  the forest and the  exponent $\beta$
depends  both  on physical  (redshift,  column  density interval)  and
instrumental (spectral resolution, decomposition of velocity profiles)
factors.

In Fig.~\ref{dndz} we  plot the result for the QSOs  in our sample for
the  standard column  density interval  $13.64 <  \log N($\HI$)  < 17$
cm$^{-2}$ in order to compare our statistics with the HST low redshift
measurement\footnote{The lower  limit in column density is  due to the
fact that HST measurements have been transformed from equivalent width
into  \HI\ column densities  assuming a  typical Doppler  parameter of
$30$ km s$^{-1}$} \citep{HST}. The  best fit to our data gives: $dn/dz
= (166\pm 4)\,  [(1+z)/3.5]^{2.8\pm 0.2}$ (1\,$\sigma$ errors).  There
is no substantial change in the trend with respect to previous results
by Kim and collaborators (2001, 2002) who used smaller samples of UVES
QSO   spectra  of  the   same  quality.    However,  our   points  are
systematically higher  on the plot,  with an increase in  $\log dn/dz$
amounting to $\sim 0.03$ at $z \sim 2$ up to $\sim 0.1$ at $z \sim 3$.
The discrepancy arises  from the fact that we  have taken into account
the decrease in the available redshift interval due to the presence of
metal  lines `masking'  the  \Lya features.   High resolution  spectra
allow to identify a larger number  of metal lines: in our sample these
metal  masks correspond  to  about  9 percent  of  the total  redshift
interval covered by the observed \Lya forests.

Fig.~\ref{dndz2} shows the number  density evolution for two different
\HI\ column  density ranges: $13 \le  \log N($\HI$) \le  14$ and $14.5
\le  \log N($\HI$) <  17\ \cmd$.   The linear  fit in  these intervals
gives slopes  of $1.9\pm 0.2$  and $3.8\pm 0.4$  for the weak  and the
strong lines selection, respectively.   This trend was already noticed
by  \citet{kim02}:  stronger  lines  have  a  steeper  number  density
evolution than the weaker ones.

\subsection{Two-point correlation function of \Lya  lines}

To study  the clustering  properties of our  sample of \Lya  lines, we
adopt the  standard two point  correlation function (TPCF)  defined as
the excess,  due to clustering, of  the probability $dP$  of finding a
\Lya  absorber  in  a volume  $dV$  at  a  distance $r$  from  another
absorber:   $dP=\Phi_{{\rm  Ly}\alpha}(z)   dV  [1+\xi   (r)]$,  where
$\Phi(z)$ is the average space  density of the absorbers as a function
of $z$.

\begin{figure}
\begin{center}
\includegraphics[width=9cm]{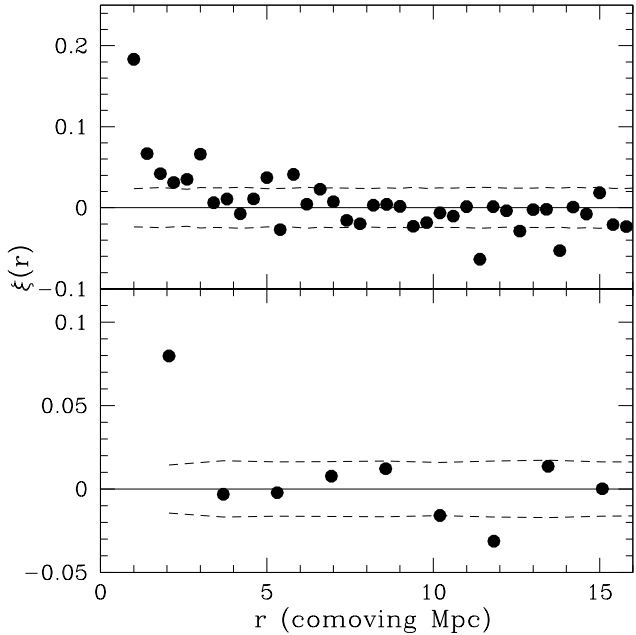}
\caption{Two point correlation function for the observed \Lya lines in
  the column  density range $12 <  \log N($\HI$) < 17\  \cmd$.  In the
  bottom panel  lines closer  than one Jeans  length have  been merged
  into one line, see text.  The dashed lines represent the $1\,\sigma$
  confidence levels from a random distribution of lines.  }
\label{TP}
\end{center}
\end{figure}

\begin{figure}
\begin{center}
\includegraphics[width=9cm]{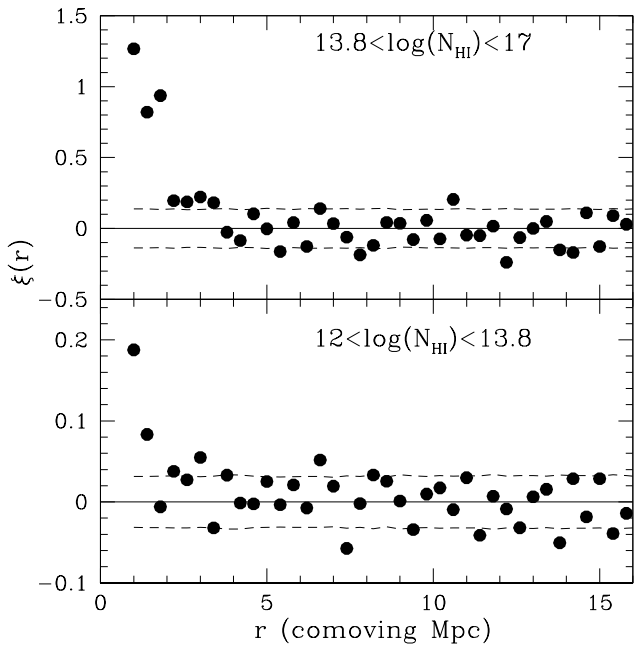}
\caption{Two point correlation function for the observed \Lya lines in
  two column density ranges as reported in the panels.}
\label{TP_cut}
\end{center}
\end{figure}

Operatively this quantity is estimated with the formula \citep{peb80}:

\begin{equation}
\xi(v)=\frac{N_{\rm obs}(v)}{N_{\rm exp}(v)}-1,
\end{equation}

\noindent
where $N_{\rm obs}$ is the observed number of line pairs with velocity
separations between $v$ and $v+dv$, and $N_{\rm exp}$ is the number of
pairs  expected  in  the  same  range of  separations  from  a  random
distribution in  redshift. Since  in this context  peculiar velocities
are negligible \citep[see e.g.][]{rauch05}, we compute the correlation
function in real space, measuring separations in comoving Mpc. At
the  characteristic  redshift  of  our  sample,  $z=2.5$,  a  velocity
separation $\Delta v=100$ \kms\  corresponds to $\Delta r\simeq 0.9$
comoving Mpc,  in our fiducial cosmology.  $N_{\rm  exp}$ is obtained
by averaging the  results of 1000 numerical simulations  of the number
of lines  observed in  each QSO spectrum.   In particular, the  set of
line redshifts is randomly generated  in the same redshift interval as
the   data   according   to   the   observed   distribution   $\propto
(1+z)^{\beta}$, where  we adopt the value  $\beta = 2.8$  found in the
previous section.  The  same mock line lists are  used to estimate the
error  on  the  observed   correlation  function  by  determining  the
$1\,\sigma$  standard deviation  of the  correlation functions  of the
randomly distributed  lines. Lines closer than  0.3 comoving Mpc,
are merged into a single line
with  redshift  equal to the  mean  redshift,  weighted  with the  column
densities,  and  column  density  equal  to  the  sum  of  the  column
densities. The minimal separation is  set by the intrinsic blending due
to the tipical width of the lines \citep[see][]{giallongo96}.

We  compute   the  correlation  function   for  the  whole   data  set
(Fig. ~\ref{TP})  and for two column  density cuts (Fig.~\ref{TP_cut})
to  investigate the clustering  properties of  strong and  weak lines.
Previous results \citep{cristiani95,lu96,cristiani97} already showed a
significant  clustering  signal   for  strong  absorptions,  which  is
confirmed and  strengthened by our  data.  Furthermore, we also  see a
significant clustering  for the  weak lines, consistent  with previous
results by \cite{misawa04}.
The  amplitude is  about one  order of  magnitude lower  than  for the
stronger  lines  but  the  clustering  signal  in  the  first  bin  is
significant at the 7~$\sigma$ level.

\begin{figure}
\begin{center}
\includegraphics[width=9cm]{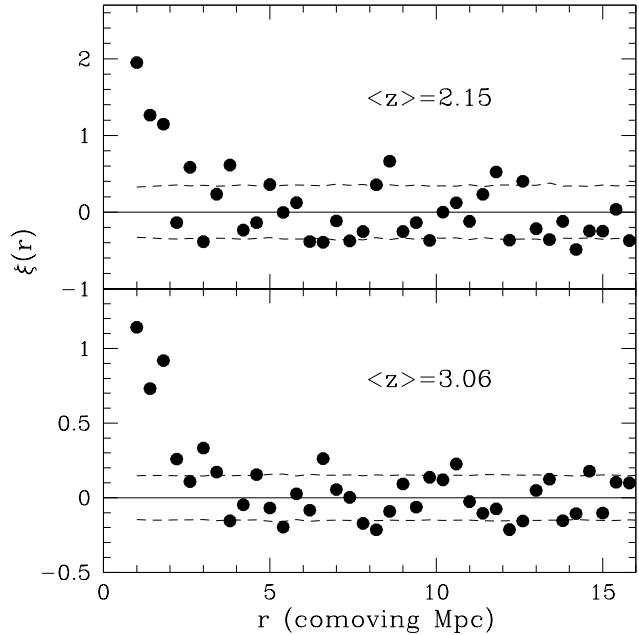}
\caption{Two point correlation function for the observed \Lya lines in
  the column  density range  $13.8 <  \log N$(\HI)$ <  17$ and  in two
  redshift ranges reported in the panels.}
\label{TP_z}
\end{center}
\end{figure}

\begin{figure}
\begin{center}
\includegraphics[width=9cm]{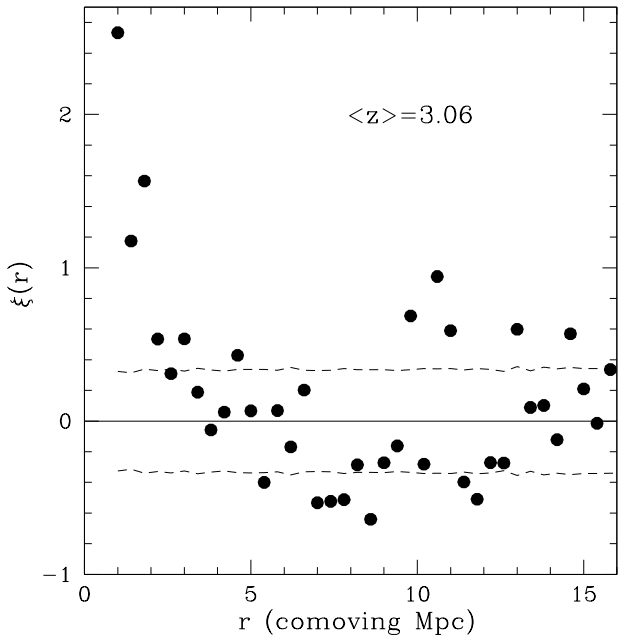}
\caption{Two point correlation function for the observed \Lya lines in
  the high-redshift range.  Here the cut in column  density is defined
  to correspond to a constant  cut in density contrast, $\delta\ga 3$,
  corresponding to $log N$(\HI)$>14.3$ at $z=3.02$, see text.}
\label{TP_z_d}
\end{center}
\end{figure}

As already said in Section~1, the Jeans length (\LJ) likely represents
the typical size of IGM  structures detected as \Lya absorptions. This
length (varying  from $\sim 1.2$ to  1.6 comoving Mpc  for the maximum
and minimum  redshift of our sample, respectively)  is also comparable
with the clustering scale of the \Lya lines as shown in Figs.~\ref{TP}
(upper  panel)  and  \ref{TP_cut}.    In  order  to  verify  that  the
clustering  signal we  are detecting  is  not only  due to  structures
internal to the  absorbers, we perform the following  test. Lines with
separation less than the local \LJ\ are merged into a single line with
column  density  equal to  the  sum of  the  column  densities of  the
component features and redshift  equal to the $N($\HI$)$-weighted mean
of the component  redshifts and the TPCF is  re-computed.  The result,
reported  in  the  lower   panel  of  Fig.~\ref{TP},  shows  that  the
clustering signal is preserved substantially  at the same level of the
one  computed   with  all  the   lines,  with  a   slightly  decreased
significance  due to the  smaller statistics.   This is  an indication
that  \Lya absorbers  cluster  among themselves  and  not only  inside
themselves.

The present data  set is large enough to  allow studying the evolution
of  the correlation  function with  redshift. We  consider  the column
density   range  for   which  the   signal  is   stronger,  $13.8<\log
N$(\HI$)<17$,  and we  divide  our  sample in  two  parts.  The  first
sub-sample is formed by  objects with emission redshift $z_{\rm em}\le
2.5$, for  which the average  \Lya forest redshift is  $\langle z_{\rm
Ly_{\alpha}}\rangle= 2.07$, and the second sub-sample has objects with
$z_{\rm em}  > 2.5$ and  $\langle z_{\rm Ly_{\alpha}}\rangle  = 3.02$.
Results are shown in Fig.~\ref{TP_z}: the high redshift lines are less
clustered than  the low redshift lines.  This  apparent evolution with
redshift  is  biased  by  the  fact  that  the  relation  $\delta-\log
N($\HI$)$ is also $z$-dependent. Indeed, the same column density range
selects objects with a lower  density contrast at higher redshift (see
eq.~\ref{tra}) explaining the lower  clustering signal. To verify this
effect,  we  selected  lines  on  the ground  of  a  constant  density
contrast, $\delta \ga 3$, which  corresponds to $\log N$(\HI$) > 13.8$
at the average  redshift of the low redshift  sub-sample, and to $\log
N$(\HI$)  > 14.3$  at the  higher average  redshift.   The correlation
function     for    the    latter     sub-sample    is     shown    in
Fig.~\ref{TP_z_d}. Selecting the same  kind of structures, there is no
longer evidence of a significant evolution with redshift.

Tab.\ref{lynum} shows a detailed budget  of the number of lines used to
compute the TPCF in all the different selections described above.

\begin{table*}
\caption{Detailed budget of the number of lines used to compute TPCFs.
The first  column refers to the  selection carried out, in  terms of emission
redshift of the considered objects. }
\label{lynum}
\begin{center}
\begin{tabular}{llcccccccc}
\hline
Selection&N$_{\rm QSO}$&\multicolumn{8}{c}{\Lya lines}\\
&&N$_{\rm tot}^{\rm a}$&N$_{\rm mask}^{\rm b}$&N$_{b}^{\rm c}$&N$_{\rm prox}^{\rm d}$&N$_{\rm comb}^{\rm e}$&N$_{\rm merg}^{\rm f}$&N$_{\rm col}^{\rm g}$&N$_{\rm fin}^{\rm h}$\\
\hline
&&&&&&&&&\\
all QSOs&22&8435&368&644&1150&1955&380&4953&1147\\
$13.8<\log N($\HI$)<17$&&&&&&&&&\\
$\Delta r>0.3$ com. Mpc&&&&&&&&&\\
&&&&&&&&&\\
all QSOs&22&8435&368&644&1150&1955&380&1319&4781\\
$12<\log N($\HI$)<13.8$&&&&&&&&&\\
$\Delta r>0.3$ com. Mpc&&&&&&&&&\\
&&&&&&&&&\\
all QSOs&22&8435&368&644&1150&1955&380&170&5930\\
$12<\log N($\HI$)<17$&&&&&&&&&\\
$\Delta r>0.3$ com. Mpc&&&&&&&&&\\
&&&&&&&&&\\
$z_{\rm em}\leq 2.5$&11&3188&103&169&445&665&100&2042&381\\
$13.8<\log N($\HI$)<17$&&&&&&&&&\\
$\Delta r>0.3$ com. Mpc&&&&&&&&&\\
&&&&&&&&&\\
$z_{\rm em}> 2.5$&11&5247&265&475&705&1290&280&2911&766\\
$13.8<\log N($\HI$)<17$&&&&&&&&&\\
$\Delta r>0.3$ com. Mpc&&&&&&&&&\\
&&&&&&&&&\\
$z_{\rm em}> 2.5$&11&5247&265&475&705&1290&280&3345&332\\
$14.3<\log N($\HI$)<17$&&&&&&&&&\\
$\Delta r>0.3$ com. Mpc&&&&&&&&&\\
&&&&&&&&&\\
all QSOs&22&8435&368&644&1150&1955&2944&64&3472\\
$12<\log N($\HI$)<17$&&&&&&&&&\\
$\Delta r>1$\LJ\ &&&&&&&&&\\
&&&&&&&&&\\
\hline
\end{tabular}
\end{center}
\begin{flushleft}
$^{\rm a}$  total number of fitted  \Lya lines; $^{\rm  b}$ number of
\Lya lines  falling in  the metal masks;  $^{\rm c}$ number  of \Lya
lines  with $b<10$  or $b>100$;  $^{\rm  d}$ number  of lines  falling
closer than  1000 \kms\  red-wards the \Lyb  emission or  closer than
5000  \kms\  blue-wards  the  \Lya  emission; $^{\rm  e}$  number  of
eliminated  lines  because  one   of  the  three  previous  conditions
occurs; $^{\rm  f}$ number of merged lines  because their separation
is  less than  the $\Delta  r$ threshold  indicated in  the selection;
$^{\rm g}$ number of merged lines not fulfilling the column
density  selection;  $^{\rm  h}$  number  of lines used to compute
the TPCF.
\end{flushleft}
\end{table*}

\section{Introducing {\small FLO}}
\label{proc}

In Section~1  we have described:  on the one  hand, what are  the main
drawbacks  of the  two  standard approaches  (Voigt  fitting and  flux
statistics) adopted to analyse  the \Lya forest and derive statistical
quantities  describing the  physical state  of the  IGM. On  the other
hand,  the recovered H  density field  is introduced  as a  new robust
estimator,  whose statistical  properties are  in good  agreement with
those  of  the  original  density  field, and  which  allows  an  easy
comparison between  observation and simulation  results.  The relation
between the underlying  H density field and the  \HI\ column densities
measured  for   the  observed   absorption  lines  is   summarised  by
eq.~\ref{tra}.

Before  describing  the  {\small  FLO}  technique in  details,  it  is
important to recall the main hypotheses:
\par\noindent 
1.  \Lya absorbers have typical sizes  of the order of the local \LJ,
which can be approximated as \citep{zaroubi}:
\begin{eqnarray}
L_{\rm J} & \simeq
&1.33\left(\frac{\Omega_{\rm m}h^{2}}{0.135}\right)^{-1/2}\left(\frac{T_0}
{1.8\times10^{4}}\right)^{1/2} \\ 
&&
\left(\frac{1.6}{\ (\alpha + 1)}\right)^{1/2}\left(\frac{1+z}{4}\right)^{-1/2} {\rm Mpc}
\nonumber 
\label{LJ}
\end{eqnarray}
in comoving units, where  $h \equiv H_{0}/100$ km s$^{-1}$ Mpc$^{-1}$,
and the other parameters have been already defined;
\par\noindent
2. the  IGM  gas  is  in  the linear  or  slightly  non-linear  regime
\citep[$\log(\delta+1)\sim 1$,][]{huignedin}


%
%

In order to  apply eq.~\ref{tra} we have, first of  all, to go through
the  Voigt fitting process  of the  \Lya forest  absorptions in  a QSO
spectrum.  Then,  to transform  the list of  \HI\ column  densities of
\Lya lines into the matter density field which generated them, we have
to perform the following steps:
\par\noindent (1)  group \Lya lines into  absorbers of size  of 1 \LJ\
with column density equal to  the sum of column densities and redshift
equal to the weighted average  of redshifts, using column densities as
weights.     
The absorbers  are created with a friend-of-friend algorithm:
\begin{enumerate} 
\item the  spatial separation between  all the possible line  pairs is
computed and the minimum separation  is compared with \LJ, computed at
the \NHI-weighted redshift mean of the pair;
\item if  the two lines  of the pair  are more distant than  the local
\LJ,  they  are classified  as  two  different  absorbers, stored  and
deleted from the line list;
\item  if the  two  lines are  closer  than the  local  \LJ, they  are
replaced in  the line list  by one line  with a redshift equal  to the
\NHI-weighted mean of the two  redshifts and a column density equal to
the sum of the two column densities;
\item the procedure is iterated until all the lines are converted into
absorbers.
\end{enumerate}
\par\noindent (2) transform the  list of column densities of absorbers
into a list of $\delta$ with eq.~\ref{tra};
\par\noindent (3)  bin the redshift  range covered by the  \Lya forest
into  steps of 1  \LJ\ and  distribute the  absorbers onto  this grid,
proportionally to  the superposition  between absorber size  (which is
again 1  \LJ) and bin.  Empty bins are  filled with one  absorber with
hydrogen  density  contrast corresponding  to  the minimum  detectable
column density in our data, $\log N($\HI$) = 12\ \cmd$ at the redshift
of the bin.
\par\noindent (4)  normalise the resulting $\delta$ field  in order to
have  $\langle \delta  +  1\rangle  = 1.0$  for  the whole  considered
sample.  This operation is necessary to recover the correct asymptotic
behaviour of the correlation function.
\begin{figure}
\begin{center}
\includegraphics[width=9cm]{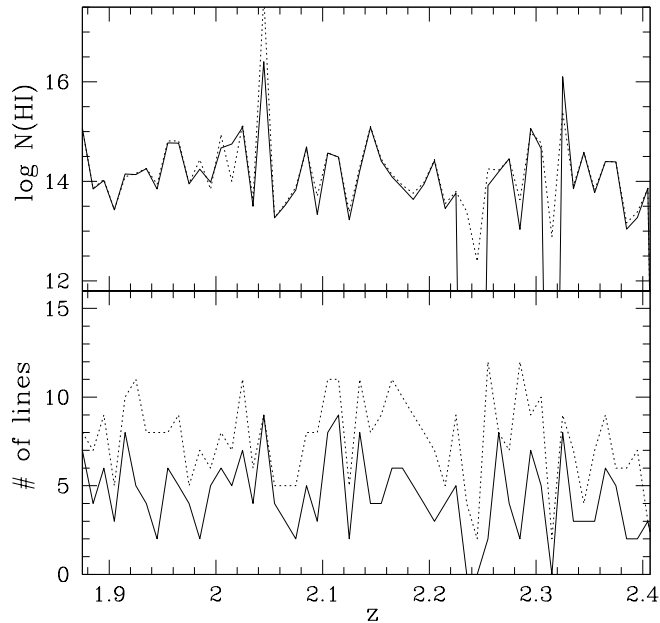}
\caption{Comparison between  the fitting results  by {\small FITLYMAN}
  (solid  line)   and  {\small  VPFIT}  (dotted  line)   for  the  QSO
  Q0109-3518.  The lower panel shows  the total number of lines, while
  the upper one shows the sum of the column densities of all the lines
  in  redshift  bins  of  width  $\Delta  z  =0.01$.The  two  redshift
  intervals where  the column density measured  with {\small FITLYMAN}
  goes  to zero  correspond to  masked  metal lines  falling at  those
  redshifts.  }
\label{fig2}
\end{center}
\end{figure}

With the introduction of this new statistical estimator, the drawbacks
of the standard Voigt  fitting approach are significantly reduced.  On
the one hand,  the statistical weight of weak  lines is reduced, since
their contribution to  the $\delta$ field is low.   On the other hand,
we verify  that, in  the process of  Voigt fitting  complex absorption
features, the total \HI\ column density is a much more robust quantity
than  the number  of  components.   To this  purpose,  we compare  the
results of  the line lists of a  sub-sample of 12 QSOs  adopted in the
present work  with the corresponding  lists obtained with  the {\small
VPFIT}  package, kindly  provided to  us  by Tae-Sun  Kim.  The  total
number of lines in each line of sight is not conserved, in particular,
significant  differences  are  observed  for  the  complex  absorption
systems where, in general, {\small VPFIT} fits more lines than {\small
FITLYMAN}.  Most of these  discrepancies are due to the identification
of  low column  density lines.  However, the  total column  density in
these complex absorbers appears to be much more stable between the two
fitting methods.

In Fig.~\ref{fig2},  we plot the comparison  between {\small FITLYMAN}
and  {\small VPFIT}  for one  QSO of  the sample,  Q0109-3518 ($z_{\rm
em}=2.407$).  We divide the line  of sight into redshift bins of width
$\Delta z=0.01$;  we sum both the  number and the  column densities of
the lines in each bin, and  plot them against redshift.  It is evident
that while  the number of lines  is different, the  two column density
distributions  trace each  other  more faithfully.   

\begin{figure}
\begin{center}
\includegraphics[width=9cm]{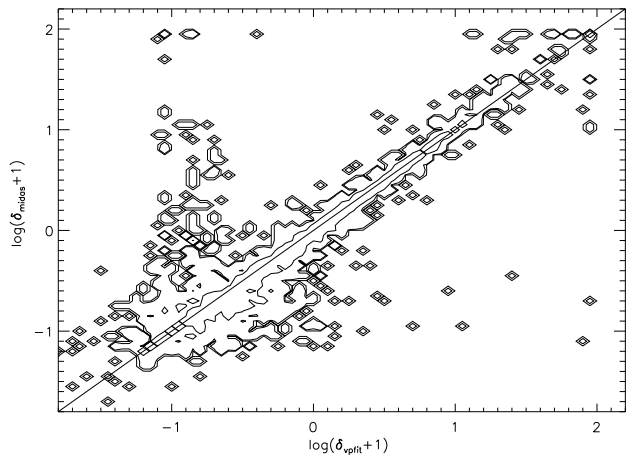}
\caption{Contour scatter plot of  the {\small FITLYMAN} versus {\small
VPFIT}  reconstructed density  fields.  The  contours show  the number
density of pixels which increases by a factor of 10 at each level.}
\label{vpmidas}
\end{center}
\end{figure}

{\small VPFIT} has been used to fit  the lines of 3 QSOs in our sample
(see  Section 2)  and  also to  analyse  the output  spectra from  the
simulation (see next section). We verify the stability of {\small FLO}
against  different fitting  tools by  applying  it to  the line  lists
obtained with  {\small VPFIT}  and with {\small  FITLYMAN} for  the 12
common QSOs.  In Fig~\ref{vpmidas}, we show the comparison between the
two  recovered  fields  by  means  of a  contour  scatter  plot.   The
correlation is tight for all values of $\delta$, the scatter increases
slightly for $\delta\lesssim 0$.

\section{Simulated data sample}
\label{sim}

We  use simulations  run with  the parallel  hydro-dynamical (TreeSPH)
code    {\small     {GADGET-2}}    based    on     the    conservative
`entropy-formulation'  of  SPH   \citep{spr05}.   They  consist  of  a
cosmological volume  with periodic boundary conditions  filled with an
equal number of dark matter  and gas particles.  Radiative cooling and
heating processes  are followed for  a primordial mix of  hydrogen and
helium.  We  assume a  mean UV-Background (UVB)  produced by  QSOs and
galaxies as given by \cite{har96} with helium heating rates multiplied
by a  factor 3.3 in order  to better fit  observational constraints on
the temperature evolution of the IGM.  This background gives naturally
a  $\Gamma \sim  10^{-12}$ (H  ionisation  rate) at  the redshifts  of
interest here \citep{bolt05}.  The  star formation criterion is a very
simple one that converts in collision-less stars all the gas particles
whose temperature falls  below $10^5$ K and whose  density contrast is
larger than 1000 (it has  been shown that the star formation criterion
has  a negligible  impact on  flux statistics).   More details  can be
found in \cite{viel04}.

We  use $2\times  400^3$  dark matter  and  gas particles  in a  $120\
h^{-1}$ comoving  Mpc box (although  for some cross-checks  we analyse
some smaller  boxes of $60\ h^{-1}$ comoving  Mpc).  The gravitational
softening  is  set  to  5  $h^{-1}$  kpc in  comoving  units  for  all
particles.

We  stress that  the  parameters chosen  here,  including the  thermal
history   of  the  IGM,   are  in   reasonably  good   agreement  with
observational  constraints including  recent  results on  the CMB  and
other    results    obtained   by    the    \Lya   forest    community
\citep[e.g.][]{viel06}.

The 120 Mpc simulation box at $z=2$ is pierced to create a set of 364
mock lines of sight covering a redshift range $\Delta z\simeq 0.11$.
For each of these lines of sight, we know the density contrast, the
temperature, and the peculiar velocity pixel by pixel.  Peculiar
velocities are small, typically less than 100 \kms, and randomly
oriented, so their contribution, e.g. to the correlation function, is
in general negligible. However, since we want to compare the result of
simulations and observations, we modify the redshifts of the density
field ($z_{\rm old}$) with the peculiar velocity field to obtain the
density field in redshift space ($z_{\rm new}$) using the formula
$v_{\rm pec}(z_{\rm old}) = c\, (z_{\rm new} - z_{\rm old})/ (1 +
(z_{\rm new} + z_{\rm old})/2)$.
We  added to  the simulated spectra  a Gaussian noise  S/N=50, in
order to reproduce the observed average S/N per pixel.  The simulated
\loss\ have been fitted with Voigt profiles using an automated version
of {\small VPFIT}.
\begin{figure}
\begin{center}
\includegraphics[width=9cm]{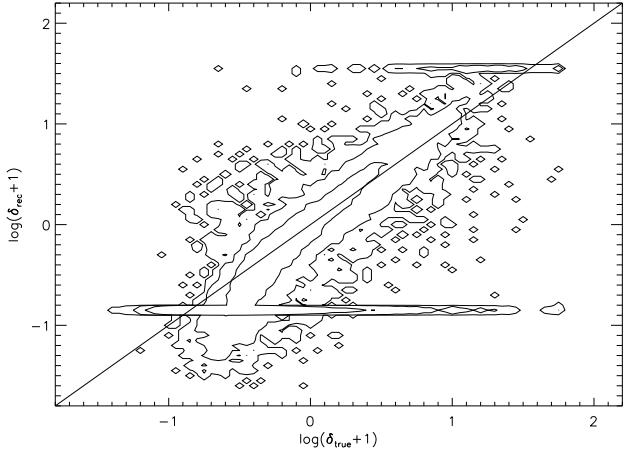}
\caption{Contour  scatter  plot   of  the  true  versus  reconstructed
  $\delta$  field  from simulations.   The  contours  show the  number
  density of pixels which increases by a factor of 10 at each level.}
\label{scatter}
\end{center}
\end{figure}
\subsection{Reconstruction of the $\delta$ field}

The \Lya lines in each simulated  line of sight are selected to have,
as  in the  case of  observations, $b  \ge 10$  \kms.  We  introduce a
further constraint,  $b \le 100$ \kms,  which is required  by the fact
that  simulated spectra are  not continuum  fitted. Shallow  and broad
oscillations in simulated spectra  are fitted as absorption lines with
Doppler  parameters of the  order of  thousands of  \kms. In  the real
spectra  these  kind  of  oscillations  are instead  fitted  with  the
continuum  and  $b$  parameters  that  large are  not  measured.   The
selected  lines are grouped  into absorbers  and transformed  into the
corresponding  density  field  following  the procedure  described  in
Section~\ref{proc}.   In  eq.~\ref{tra}  the values:  $T_0=1.8  \times
10^4$ K and  $\alpha = 0.6$ are adopted, which  are consistent with an
early  re-ionisation  epoch  and   are  the  ones  inferred  from  the
simulations.

The reconstructed  field is compared  with the original  density field
(i.e. the output of the simulation),  which is also binned into 1 \LJ\
steps.   An upper  threshold  is adopted  both  for the  true and  the
recovered $\delta$  field, $\delta_{\rm thr}=50$,  since 99.95 percent
of  pixels in the  simulated lines  of sight  have values  $\delta \le
\delta_{\rm  thr}$ and  the algorithm  to recover  the  $\delta$ field
(eq.~\ref{tra}) is  valid for values  of $\delta \le {\rm  few} \times
10$. The upper cut is applied before the normalisation process.

The  average values  of the  $\delta$  field considering  all the  364
simulated spectra  are: $\langle\delta + 1\rangle \simeq  0.9$ and 1.3
for  the  true and  recovered  field,  respectively.   The fields  are
normalised using these values.

Figure~\ref{scatter} shows  the contour  scatter plot of  the original
versus reconstructed density  field.  As can be seen  from the figure,
{\small  FLO} reconstructs fairly  well the  original field  above the
mean  density,  while   under-densities  are  underestimated,  or  not
recovered, if  they are below  our lower threshold. Indeed,  the lower
horizontal tail observed  in the scatter plot is  due to the treatment
of the empty bins  during the absorber-field transformation. The upper
horizontal tail  is instead due  to the cut applied  to over-densities
larger than the threshold, $\delta_{\rm thr}$.
\begin{figure}
\begin{center}
\includegraphics[width=9cm]{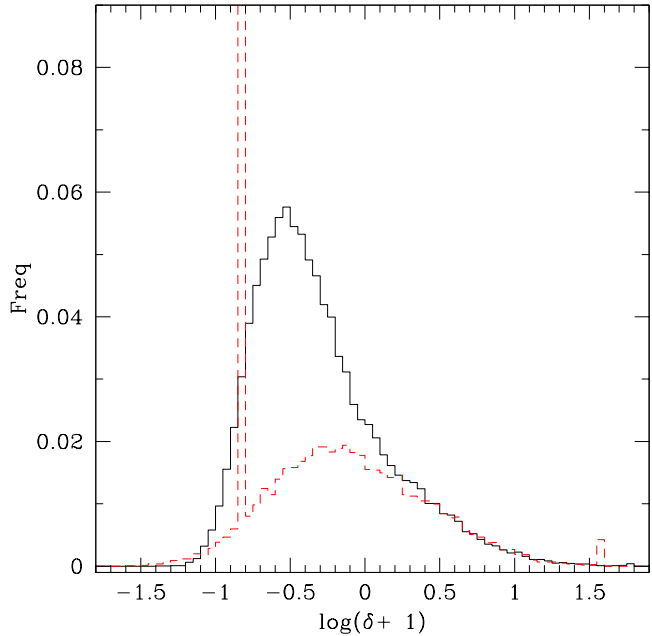}
\caption{Distribution of $\delta$ values  in the true (solid line) and
recovered  (dashed  line) field  normalised  to  the  total number  of
points.}
\label{distrib}
\end{center}
\end{figure}
Figure~\ref{distrib} shows the distribution  of $\delta$ values in the
true and recovered field. The  peak at $\log (\delta +1) \simeq -0.83$
contains $\simeq 53$\  percent of all the points and it  is due to the
procedure  that   assigns  to  empty   bins  the  value   of  $\delta$
corresponding to the  redshift of the bin and  to the minimum observed
column  density, $\log  N($\HI$)=12\ \cmd$.   On the  other  hand, the
small bump  at $\log (\delta+1) \simeq  1.57$ is due to  the upper cut
applied  to the  recovered density  field and  it  includes $\sim$~0.4
percent of the total number of points.

The transformation  starts to  recover more than  half of  the correct
values of $\delta$ at $\log (\delta +1) \simeq -0.15$ and recovers all
the $\delta$  within 30 percent in  the range $-0.08  \la \log (\delta
+1)  \la  1.45$.  Clearly,  we  are  not  dealing correctly  with  the
under-dense  regions,  even  if   they  are  above  our  observational
detection  limit.  This is  likely due  to the  fact that  our primary
hypothesis, the local hydrostatic  equilibrium, is not valid for those
regions. This was  also discussed by \citet{Sch} and  here we have the
evidence that under the mean density the gas is still expanding.

In the  next section, it  will be shown  that the inaccuracies  in the
under-density regime do  not significantly affect statistical measures
like the correlation function.

\begin{figure}								
\begin{center}								
\includegraphics[width=9cm]{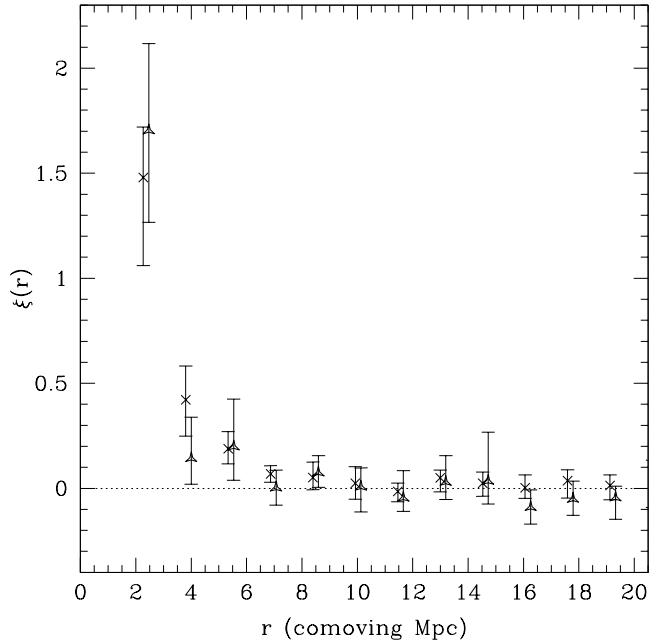}			
\caption{Correlation function  of $\delta$ from  simulations.  Crosses
  represent the correlation function  obtained from the original data,
  triangles the one obtained from the reconstructed field.}
\label{corsym}								
\end{center}								
\end{figure}                                                            

\subsection{Two-point statistics of the $\delta$ field}
\label{corsimpar}

We computed  the correlation function  for the original  and recovered
$\delta$ field, with the formula:

\begin{equation}
\label{corflu}
\xi_{\delta}( r)=\langle\delta(r+dr)\delta(r)\rangle,
\end{equation}

\noindent
where $r$  is the physical separation  of two points  in comoving Mpc.
$\xi_{\delta}(r)$   quantifies  the   clustering  properties   of  the
considered field,  showing a signal significantly  different from zero
at  separations   where  the   field  presents  structures   (over  or
under-densities).

The bin  size is  the largest  value of \LJ\  for our  sample, $\simeq
1.532$ comoving Mpc, corresponding to the minimum redshift.

Figure~\ref{corsym} plots the results  of the correlation function for
the true and  recovered $\delta$ field.  The value in  each bin is the
median  value of  50  sample of  88  lines of  sight  obtained with  a
bootstrap technique from  the 364 lines of sight  of the total sample.
This procedure  is required in order  to compare this  result with the
analogous  one for  the observed  data (see  Section~\ref{cf_obs}). We
have  22  observed  spectra  but  each one  covers  a  redshift  range
corresponding   to  about   4  simulated   spectra.    Error-bars  are
$1\,\sigma$,  computed with  the  percentiles of  the distribution  of
values in  each bin.   The recovered correlation  function is  in very
good agreement with the true one at every separation.

\begin{figure}
\begin{center}
\includegraphics[width=9cm]{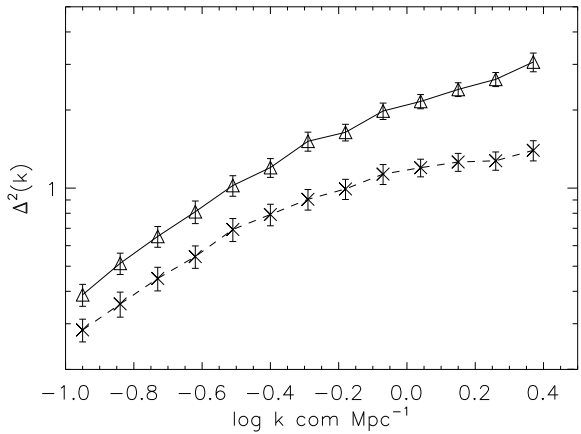}
\caption{One  dimensional  power  spectrum  of  the  hydrogen  density
  contrast field. The solid and the dashed lines represent the results
  for   the   reconstructed   and   the   original   $\delta$   field,
  respectively.}
\label{fig:pksim}
\end{center}
\end{figure}
 
We also  estimate the one  dimensional power spectrum of  the hydrogen
density contrast  field, that is  defined by the Fourier  transform of
the correlation function:

\begin{align}
&\xi(r)=\frac{V}{2\pi}\int|\delta_k|^2e^{-ikr}dk\\\nonumber
&P^{1D}_{\delta}(k)\equiv\langle|\delta_k|^2  \rangle
\label{pk}
\end{align}
The  power spectrum is  computed adopting  the Fast  Fourier Transform
(FFT) technique  which requires  that the field  to be  transformed is
evenly sampled. To this purpose,  we have re-binned the observed lines
of sight to a constant step  equal to the minimum Jeans length for the
considered \Lya forest (corresponding to the maximum redshift).  Then,
the following steps have been applied:
\par\noindent 
1) a grid of wave-numbers is built in the Fourier space, starting from
$k_{\rm min}=2\pi/\Delta\, r$,  where $\Delta\, r$ is the  length of a
\los\ in  comoving Mpc,  and formed by  $n_{\rm pix}/2$  evenly spaced
elements, where $n_{\rm pix}$ is  the number of pixels of the original
$\delta$ field;
\par\noindent 
2) the Fourier transform of the $\delta$ field is computed;
\par\noindent 
3) the products $\delta_k\delta_{k^{'}}$ are averaged in each bin;
\par\noindent  
4) the $P^{\rm  1D}_{\delta}(k)$ is  normalised by multiplying  it for
the \los~ length, $\Delta r$;
\par\noindent  
5) the obtained $P^{\rm 1D}_{\delta}(k)$ is smoothed on larger bins to
reduce the noise;
\par\noindent 
6) the result is averaged over all the lines of sight.
\par\noindent
Error bars are computed using a jackknife estimator \citep{bradley} on
the whole sample of simulated lines of sight.

Fig.~\ref{fig:pksim}  shows  the results  of  the  computation of  the
corresponding $\Delta^2(k)=kP^{\rm  1D}_{\delta}(k)/2\pi$ for the true
and recovered  simulated $\delta$ fields.   The two power  spectra are
consistent  at the  $3\,\sigma$  level on  scales  $20 \la  r \la  60$
comoving Mpc.
\begin{figure}
\begin{center}
\includegraphics[width=9cm]{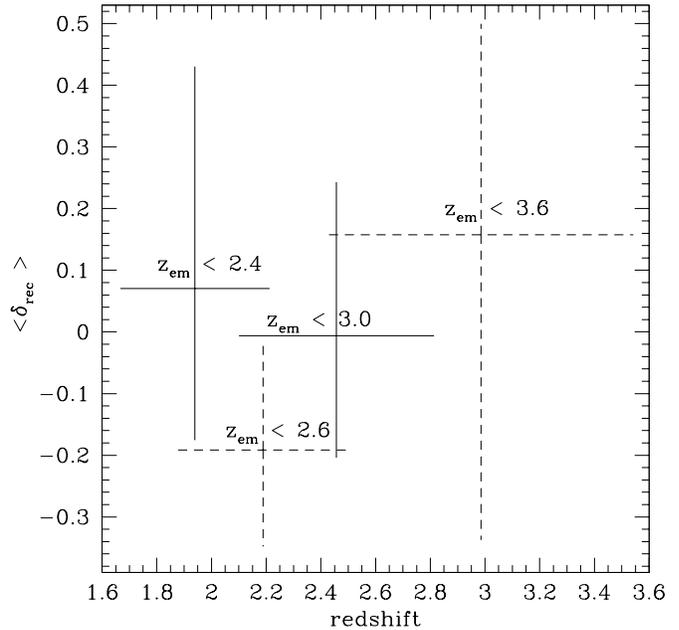}
\caption{Distribution  with  redshift  of  the  average  reconstructed
  $\delta$ values for 4 sub-samples of QSOs selected by their emission
  redshift as displayed in the plot  (formed by 6,6,5 and 5 objects as
  redshift  increases).   Horizontal  lines  represent   the  redshift
  coverage of each QSO sample,  while vertical lines are the spread in
  average $\delta$ values for the single QSOs in the samples. }
\label{nhdistr}
\end{center}
\end{figure}

\section{{\small FLO} applied to the observed data sample}

In Section~\ref{subsec:line}, we have described how the line lists are
compiled for  the 22 high  resolution QSO spectra forming  our sample.
The  procedure  explained in  Section~\ref{proc}  is  then applied  to
obtain  the corresponding  density  contrast field  for  each line  of
sight.  In the case of observations,  we have to take into account the
presence of  the masked intervals  covering regions occupied  by metal
absorption systems. We eliminate all the bins  that are covered
by  more than  30\% by  masked intervals. Before  the normalisation
step,  we apply  an  upper threshold  as  in the  case of  simulations
($\delta_{\rm thr} = 50$) since we want to compare our result with the
one  obtained   in  Section~4.2.   The  pixels   above  the  threshold
correspond  to $\sim  0.9$  percent  of the  total  number of  pixels.
Figures~\ref{vpmidas}   and  \ref{scatter}   show   the  uncertainties
associated with  the use of  a different fitting tool  (in particular,
{\small  VPFIT}  and {\small  FITLYMAN})  and  the  ones intrinsic  to
{\small FLO}, respectively.  Since the intrinsic errors turn out to be
larger  than those  induced by  the fitting  technique, we  can safely
compare the  results obtained from the simulations  and those obtained
from the data sample, presented in this section.
    
Figure~\ref{nhdistr}  shows  the   average  values  of  the  recovered
$\delta$ fields for 4 sub-samples  built from the 22 observed lines of
sight selected on the ground of the QSO emission redshifts.
The spread of average $\delta$  values for the single QSOs forming the
samples is also  shown.  There is no significant  trend with redshift,
as  it  is  expected if  the  density  field  follows on  average  the
evolution of the cosmic mean value.

\subsection{Two point statistics of the $\delta$ field}
\label{cf_obs}

The correlation  function for the observed $\delta$  field is computed
with the formula given in eq.~\ref{corflu} as for simulations.

\begin{figure}
\includegraphics[width=9cm]{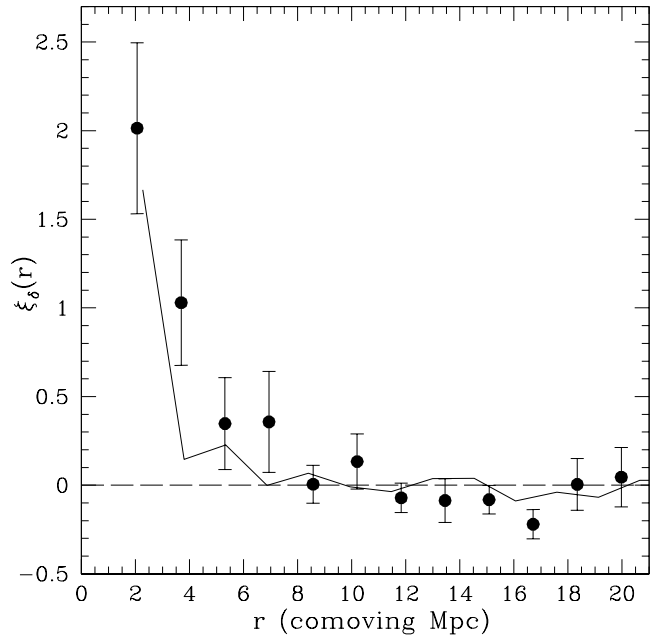}
\caption{  Correlation function  of the  $\delta$  field reconstructed
  from our  22 observed  QSO spectra.  Points  refer to the  data, the
  line instead represent the prediction from the simulation.}
\label{corrnh}
\end{figure}
 
The result is  shown in Fig.~\ref{corrnh}. Here the  value in each bin
is  obtained  averaging all  the  sample,  while  the error  bars  are
computed creating  50 samples of  22 lines of  sight drawn out  of our
sample with  a bootstrap technique  and taking the percentiles  of the
distribution corresponding to $1\,\sigma$ errors.
The bin  size is $\simeq  1.6$ comoving Mpc  which is the \LJ\  at the
lowest redshift  of the sample.  The clustering  signal is significant
at more than the $3\,\sigma$ level  in the first two bins ($r \la 4.5$
comoving  Mpc). We  have  superimposed  to the  data  points the  TPCF
obtained  from the recovered  $\delta$ field  of simulations.  The two
correlation  functions  are in  very  good  agreement, confirming  the
validity of the cosmological parameters adopted in the simulation.


\begin{figure}
\includegraphics[width=9cm]{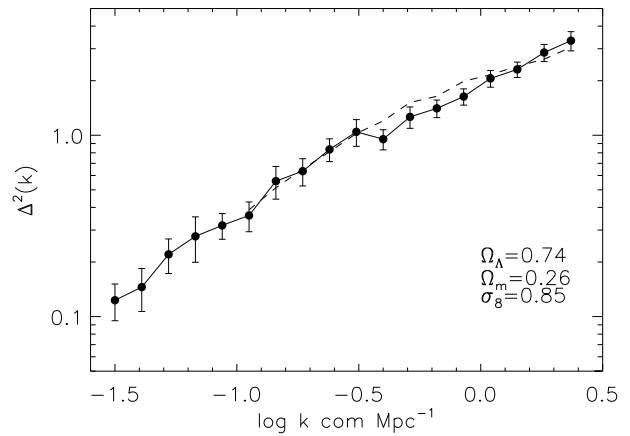}
\caption{Power spectrum  of the $\delta$ field  reconstructed from our
22 observed  QSO spectra.  The  dashed line represents  the prediction
from the simulation.}
\label{pkoss}
\end{figure}



Since  $P^{\rm  1D}_{\delta}(k)$  is  very sensitive  to  cosmological
parameters, it is very important to  check if the prediction of such a
function are in agreement with the observed values.

In the case  of the observed spectra, the masked  metal lines make the
starting  grid of  pixels unevenly  spaced,  thus not  fitted for  the
application  of the  FFT. To  overcome this  problem, as  a  1st order
approximation, the masked  bins have been put to  the average density.
This procedure  is based on the  observation that the  \Lya forest gas
traces  on average  the average  density and  on the  analogous method
adopted in  the computation of  the power spectrum of  the transmitted
flux  \citep{viel04}.   Fig.~\ref{pkoss}  shows  the  result  of  this
computation: the power spectrum obtained from our data is in excellent
agreement with the one  obtained from the density fields reconstructed
with {\small  FLO} from the  simulated spectra based on  a concordance
cosmological  model.   Error  bars  are  computed  using  a  jackknife
estimator on the whole sample of observed QSOs.  The quantity which is
generally  compared  with  the  model  predictions  is  the  3D  power
spectrum, which  is obtained from  the 1D by  differentiation, $P^{\rm
3D}_{\delta}(k)=-\frac{2\pi}{k}\frac{dP^{\rm 1D}_{\delta}(k)}{dk}$.  A
comparison of the rough estimates of $P^{\rm 3D}_{\delta}(k)$ from our
observed  and  simulated  data  gives  good  agreement.   However,  we
postpone a careful study of this quantity obtaining constraints on the
cosmlogical parameters to a forthcoming paper.
\section{Conclusions}

We have presented  results from the analysis of  the largest sample of
fitted  \Lya  lines obtained  from  22  high  resolution QSO  spectra
covering the redshift range between $\sim$ 1.7 and 3.5.

In particular, we have computed: 
\begin{itemize}
\item  [1.a)] the  line number  density evolution  with  redshift: for
  which we find $dn/dz \simeq (166\pm 4)[(1+z)/3.5]^{2.8\pm0.2}$. 
  While the  redshift evolution is consistent 
  with  previous results,  the  normalisation is  higher  by a  factor
  ranging from $\sim  0.03$ in $\log(dn/dz)$ at $z\sim  2$ to $0.1$ at
  $z\sim 3$. This  difference is due to the  improved treatment of the
  contamination by metal  lines (amounting to $\sim 9$  percent of the
  redshift  interval  covered by  the  \Lya  forests),  which is  made
  possible  by the high  resolution and  signal-to-noise ratio  of our
  spectra.  Consistently  with \citet{kim02},  we also find  a steeper
  evolution for  the stronger  lines ($14.5 \le  \log N($\HI$)  < 17$)
  compared to the weak ones ($13 \le \log N($\HI$) < 14$).
\item [1.b)] the two-point  correlation function (TPCF): which shows a
  significant clustering signal up to $\sim 2$ comoving Mpc for strong
  lines ($13.8 \le \log N($\HI$) < 17$), and
also  for weak  lines  ($12 \le  \log  N($\HI$) <  13.8$) although  on
smaller scales, $\la 1.5$ Mpc.
  We then calculated the TPCF by
  grouping  all the  lines closer  than  the local  Jeans length  (the
  assumed typical  size for  the hydrogen absorbers in the IGM). The
  signal is still significant in the first bin (r $\la$ 2.5 Mpc).
\item[1.c)]  the  TPCF  evolution   with  redshift  for  strong
  lines: we divided our sample  in two sub-samples; the first one,
  formed by objects with $z_{\rm em}\la 2.5$, for which the average
  \Lya forest redshift is $\langle  z_{\rm Ly_{\alpha}}\rangle =
  2.07$, and the second one, formed by  objects  with  $z_{\rm
  em}>2.5$, with  $\langle z_{\rm Ly_{\alpha}}\rangle = 3.02$. 
  The TPCF computed for lines with $13.8 < \log N($\HI$) < 17$ in
  these two samples, show a trend of increasing clustering with
  decreasing redshift; this  is an apparent evolution, due to the fact
  that the relation $\delta-\log(N($\HI$))$ is $z$-dependent. Indeed a
  selection of  lines tracing the same kind of structures
  (characterised by $\delta>3$) shows no evidence of a 
  significant evolution with redshift of the TPCF.
\end{itemize}

In the second part of the paper, we have described {\small FLO}, a new
algorithm to transform the measured \HI\ column densities of the \Lya
lines  detected along a  line of  sight, into  the underlying  total H
density field (and in particular, the density contrast, $\delta \equiv
n_{\rm H} /\langle n_{\rm H}\rangle  -1$, field).  The method is based
on  the  assumption that  \Lya  absorbers  are  in local  hydrostatic
equilibrium  and, as a  consequence, the  Jeans length  corresponds to
their characteristic size.  The aim of  this study is to find a robust
statistical  estimator which  allows  a direct  link  to the  physical
properties  of the  gas and  an easy  comparison with  the  results of
simulations.  To test the effects  of the transformation, we have used
a  set   of  364  lines  of   sight  obtained  from   a  large  N-body
hydro-dynamical  simulation run  in  a box  of  $120 h^{-1}$  comoving
Mpc. For  every line of  sight we have  both the density  and velocity
field pixel  per pixel and the  list of Voigt fitted  \Lya lines with
central redshift,  column density and Doppler  parameter.  Our results
can be summarised as follows:

\begin{itemize}
\item [2.a)] {\small FLO}  recovers extremely well (within 30 percent)
  the  over-densities  up  to  $\delta   \sim  30$  while  it  is  not
  reproducing correctly  the under-densities (more than  50 percent of
  $\delta$  values are  not recovered)  even  in the  range above  our
  resolution  limit.   This result  suggests  that  the hypothesis  of
  hydrostatic  equilibrium is  not valid  for the  under-dense regions
  that are likely still expanding. On  the other hand, for the goal of
  our study, that is the  computation of statistical properties of the
  IGM, the  resulting $\delta$  field gives satisfactory  results when
  the  two-point correlation function  and the  1D power  spectrum are
  considered.  The  comparison of the  results obtained with  the true
  $\delta$ field of the simulation and with the one reconstructed from
  line  column  densities  with   our  algorithm,  are  in  very  good
  agreement.

\end{itemize}
\par\noindent When  applied to the  observed data sample,  the {\small
  FLO} technique gives the following results:
\begin{itemize}
\item [2.b)] the redshift distribution of the average hydrogen density
  is consistent with the evolution of the cosmic mean hydrogen density
  in the redshift range covered by our QSO sample, supporting the fact
  that the \Lya  forest arises from fluctuations of  the IGM close to
  the mean density;
\item [2.c)]  the correlation function  of the density  field obtained
from the observed spectra shows  a significant clustering signal up to
$\sim  4$ comoving  Mpc and  is consistent  with the  analogous result
obtained for the recovered density  field in a simulation based on the
concordance cosmological model.
\item [2.d)] the one dimensional  power spectrum of the $\delta$ field
obtained from the observed spectra  is in very good agreement with the
same  result  obtained  from  the  recovered density  field  from  the
simulation  based  on  the  concordance cosmological  model  on  scale
lengths between $\sim 2.5$ and 63 comoving Mpc.
\end{itemize}

The algorithm presented in this work is particularly useful to extract
information  from observations  in terms  of overdensities,  making it
possible a more direct and handy comparison with simulations.

\section*{Acknowledgments}                        
Hydro-dynamical  simulations were  done at  the UK  National Cosmology
Supercomputer  Center funded by  PPARC, HEFCE  and Silicon  Graphics /
Cray  Research and at  the HPCF  (Cambridge High  Performance Computer
Cluster).  We are grateful to  Tae-Sun Kim for permission of using her
lists of  fitted QSO absorption lines before  publication.  HIRES data
were obtained at the W.M. Keck Observatory, which was made possible by
the  generous financial  support  of the  W.M.   Keck Foundation,  and
operated  as a  scientific partnership  among the  California  and the
National Aeronautics and Space Administration.

\label{lastpage}
\end{document}